\documentclass[twocolumn,showpacs,aps,superscriptaddress,preprintnumbers,floatfix,amsmath,amssymb,prd]{revtex4-1}

\usepackage{dcolumn}% Align table columns on decimal point
\usepackage{bm}% bold math
\usepackage{graphicx}% Include figure files

\newcommand{\ve}[1]{\ensuremath{\mathbf{#1}}}
\newcommand{\n}[1]{\ensuremath{|\mathbf{#1}|}}
\newcommand{\qrec}{\ensuremath{Q_{\textrm{rec}}^2}}
\newcommand{\erec}{\ensuremath{E_\nu^{\textrm{rec}}}}

\newcommand{\EB}{\ensuremath{\varepsilon}}
\def\lsim{\lesssim}

\def\beq{\begin{equation}}
\def\eeq{\end{equation}}
\def\be{\begin{eqnarray}}
\def\ee{\end{eqnarray}}

\begin{document}

\title{Analysis of the $Q^2$ dependence of charged-current quasielastic processes\\ in neutrino-nucleus interactions}

\author{Artur M. Ankowski}
\email{Artur.Ankowski@roma1.infn.it}
\affiliation{Institute of Theoretical Physics, University of Wroc{\l}aw,\\ Wroc{\l}aw, Poland,\footnote{Permanent address.}}
\affiliation{INFN and Department of Physics,``Sapienza'' Universit\`a di Roma,\\ I-00185 Roma, Italy}
\author{Omar Benhar}
\affiliation{INFN and Department of Physics,``Sapienza'' Universit\`a di Roma,\\ I-00185 Roma, Italy}
\author{Nicola Farina}
\affiliation{INFN and Department of Physics,``Sapienza'' Universit\`a di Roma,\\ I-00185 Roma, Italy}

\date{\today}

\graphicspath{{plot/}}

\begin{abstract}
We discuss the observed disagreement between the $Q^2$ distributions of neutrino-nucleus
quasielastic events, measured by a number of recent experiments, and the predictions of
Monte Carlo simulations based on the relativistic Fermi gas model. The results of our analysis
suggest that  these discrepancies are likely to be ascribable to both the breakdown of the impulse approximation
and the limitations of the Fermi gas description. Several issues related to the extraction of the $Q^2$ distributions
from the experimental data are also discussed, and new kinematical variables, which would allow for an improved
analysis, are proposed.

\end{abstract}

\pacs{13.15.+g, 25.30.Pt}

%\keywords{Suggested keywords}%Use showkeys class option if keyword display desired
\maketitle

\section{Introduction}

A number of recent experiments reported a sizable disagreement between the measured $Q^2$ distributions
of neutrino-nucleus scattering events and the results of Monte Carlo (MC) simulations \cite{ref:K2K,ref:MiniB_eff,ref:SciBooNE}. The largest discrepancies
occur in the region of low $Q^2$, typically $Q^2 \lsim  0.2$ GeV$^2$, where the observed number of charged-current
quasielastic (CCQE) events is significantly lower than MC predictions.
To account for this differences, the MiniBooNE Collaboration introduced an
additional parameter in the relativistic Fermi gas (RFG) model used in the MC code~\cite{ref:MiniB_eff}. This procedure
leads to a modification of the treatment of Pauli blocking, whose physical interpretation is quite questionable. In
Ref.~\cite{ref:MiniB_eff} it was also argued that a more refined treatment of nuclear effects may be needed.
However, the results of Ref.~\cite{ref:axialMass} suggest that replacing the RFG model with the approach based on the use of a realistic spectral
function (SF) does not lead to a consistent description of low- and high-$Q^2$ data. Note that, although strictly speaking the
spectral function can also be defined in the RFG model, in which case it reduces to a collection of $\delta$-function peaks,
in the following we will use the acronym SF to indicate a spectral function obtained from more advanced dynamical models.

The basic assumption underlying MC simulations is the validity of the impulse approximation (IA),
implying that the scattering process involves only one nucleon, while the remaining $(A-1)$ particles act as
spectators. This scheme is likely to be applicable if the space resolution of the incoming neutrino,
$\lambda \sim {\n q}^{-1}$, where $\ve q$ is the momentum transfer, is small compared to the average
separation between nucleons in the target nucleus.

In this paper we analyze possible sources of the observed discrepancies, associated with both the breakdown of the IA
and the limitations of the RFG model, and point out some issues related to the extraction
of the $Q^2$ distributions from the experimental data.

In Section \ref{sec:Q2} we discuss the bias of the $Q^2$ reconstruction from the kinematics of the CCQE events, while Section
 \ref{lowq} is devoted to a quantitative investigation of the limits of applicability of the IA scheme.
In Section \ref{expt:issues} we discuss the main difficulties involved in the comparison between theoretical calculations
and experimental data, and put forth the proposal of new kinematical variables,  whose use may allow for a more effective
data  analysis. Finally, in Section \ref{concl}, we summarize our results and state the conclusions.

\section{Uncertainties in the $Q^2$ reconstruction}
\label{sec:Q2}

%\subsection{Reconstructed vs. theoretical $Q^2$}

Consider the yield of CCQE events averaged over the MiniBooNE neutrino flux. We have
compared the distributions of events plotted as a function of $Q^2$ and its reconstructed value, defined as
\cite{ref:K2K,ref:MiniB_eff}
%%%%%%%%%%%%%%%
\begin{equation}\label{eq:recQ2}
\qrec=-m_\mu^2+2\erec(E_\mu-\n{k'}\cos\theta),
\end{equation}
%%%%%%%%%%%%%%%
where $m_\mu$, ${\bf k}^\prime$, and $\theta$ denote the muon mass, momentum, and scattering angle, respectively.
In the above equation
%%%%%%%%%%%%%%%
\begin{equation}\label{eq:recE}
\erec= \frac{2E_\mu(M_n - \EB)-(\EB^2 - 2 M_n\EB + m_\mu^2 + \Delta M^2)}{2 ( M_n - \EB - E_\mu + \n{k'}\cos\theta )},
\end{equation}
%%%%%%%%%%%%%%%
with $\Delta M^2=M_n^2 - M_p^2$,  $M_n$ and $M_p$ being the neutron and proton mass, is the reconstructed energy of the incoming neutrino.
Note that Eqs.~\eqref{eq:recQ2} and \eqref{eq:recE} have to be regarded as definitions of  two quantities
used in data analysis and should not be identified with the true $Q^2$ and neutrino energy.

\begin{figure}[t]
    \centering
    \includegraphics[width=0.80\columnwidth]{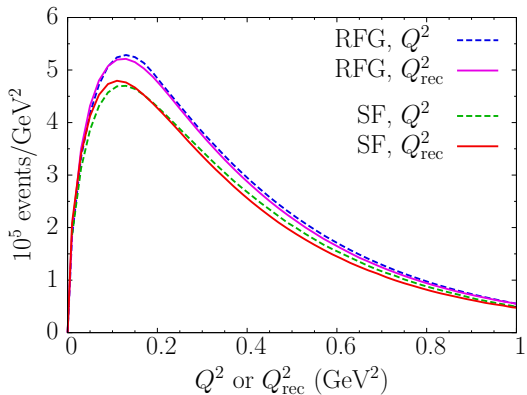}%
\caption{\label{fig:Q2vsQ2rec} (color online). Distributions of CCQE events in carbon as a function of $Q^2$ and $\qrec$. The calculations
have been carried out for the MiniBooNE flux, using the RFG model (upper curves) and the SF approach (lower curves).}
\end{figure}

The results of the calculations, carried out for a carbon target using both
the RFG model, with Fermi momentum $p_F=220$ MeV and separation energy $\EB=34$ MeV, and the SF approach,
with the spectral function of Ref.~\cite{SFCarbon}, are shown in Fig.~\ref{fig:Q2vsQ2rec}. In the SF approach
Pauli blocking has been taken into account using realistic nuclear matter momentum distributions and the local density
approximation, within the
scheme discussed in Section~\ref{lowq}.
%consistently taken into account using the momentum distribution obtained through energy integration of
%the hole spectral function (see the discussion in Section~\ref{lowq}).
It is apparent that in both cases the definition of $\qrec$ in terms of the measured quantities $\n{k'}$ and $\theta$ allows one to reproduce quite well the results obtained using the true $Q^2$. However, as the results shown in Fig.~\ref{fig:Q2vsQ2rec} involve a flux average, one may
still ask the question whether the two quantities, $Q^2$ and $\qrec$, are totally equivalent.
%%%%%%%%%%%%%%%%
%%%%%%%%%%%%%%%%
%%%%%%%%%%%%%%%%
\begin{figure}
    \centering
    \includegraphics[width=0.80\columnwidth]{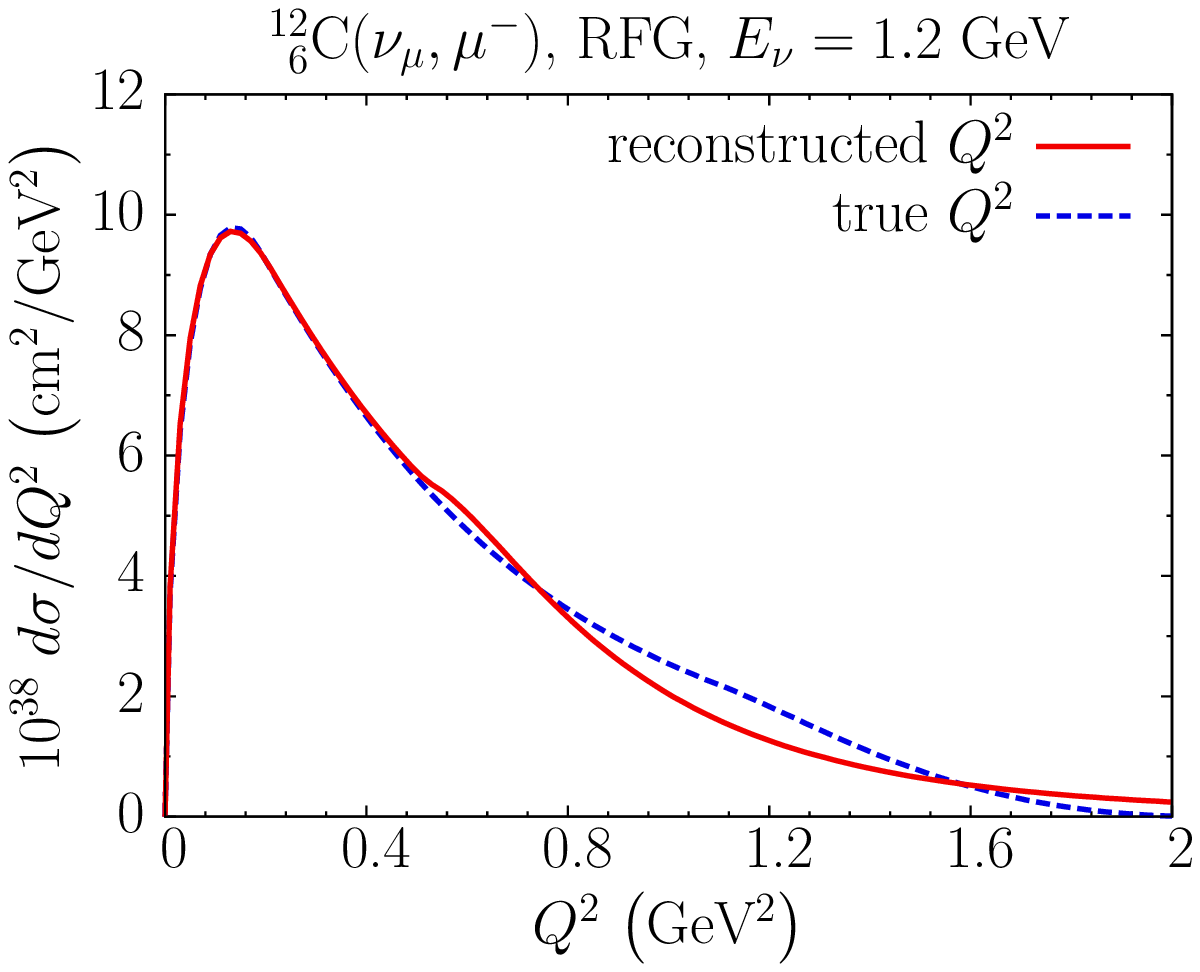}\\
    \vspace{1em}
    \includegraphics[width=0.80\columnwidth]{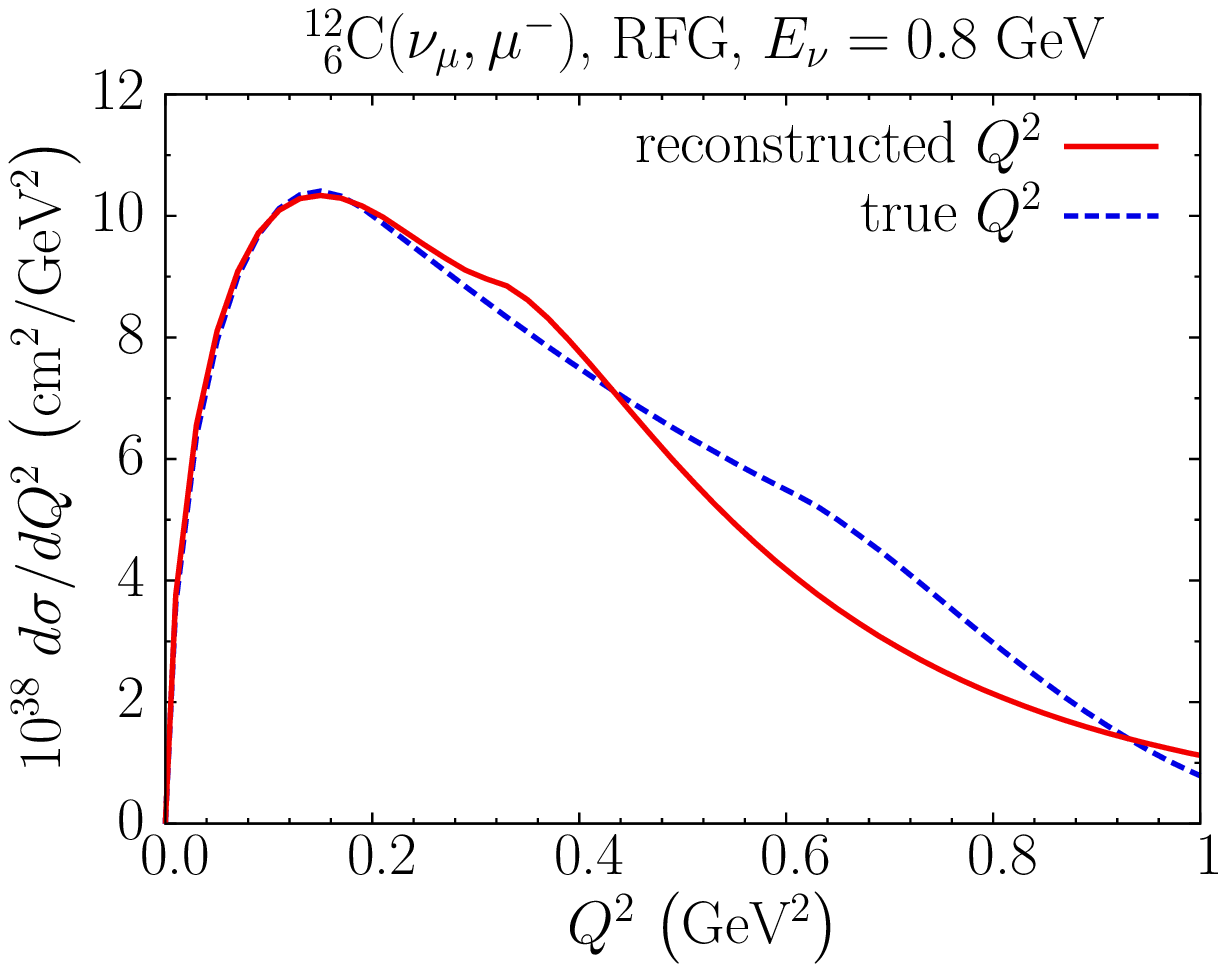}\\
    \vspace{1em}
    \includegraphics[width=0.80\columnwidth]{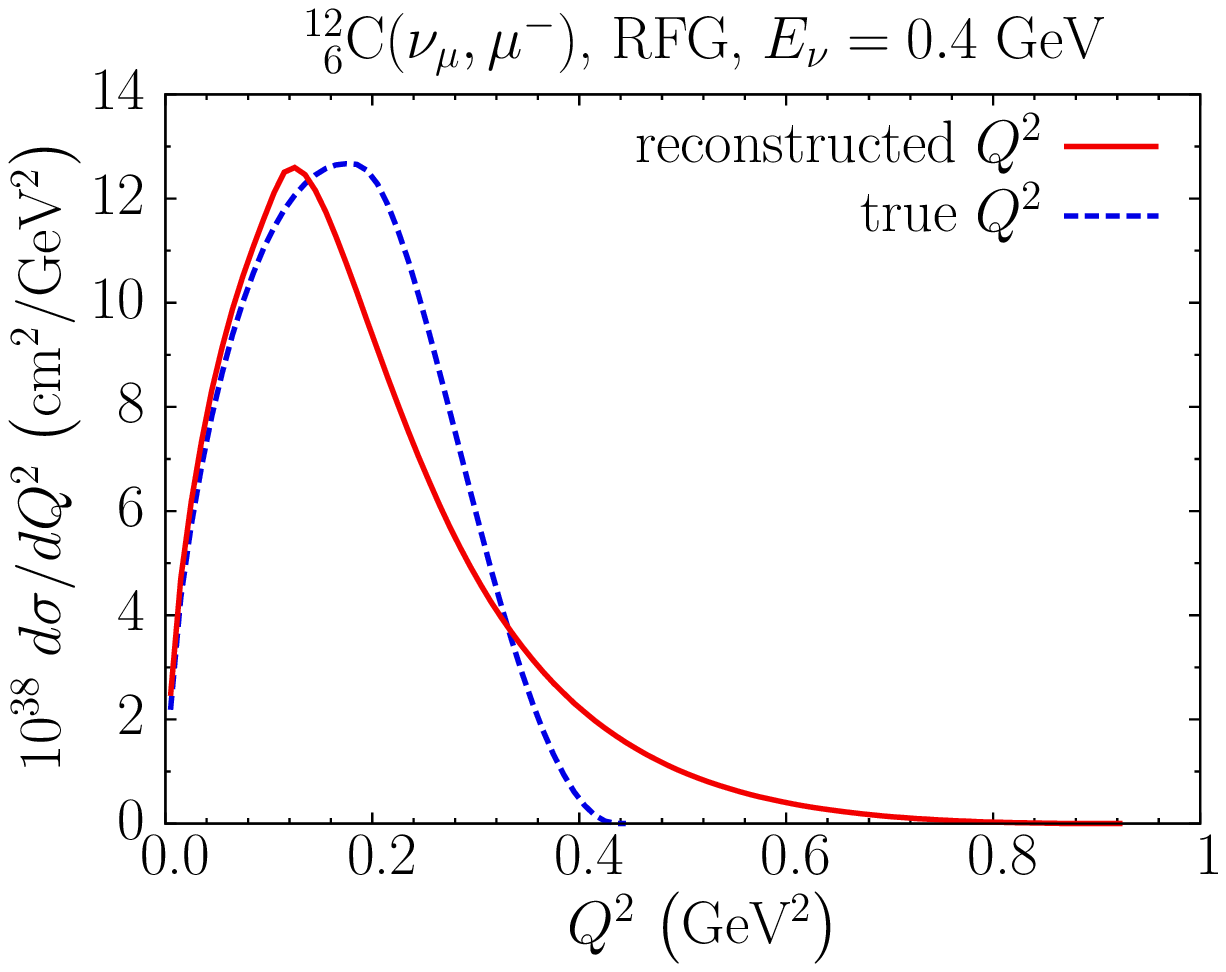}
\caption{\label{fig:givenE} (color online). Comparison of the differential cross sections $d\sigma/dQ^2$ and $d\sigma/d\qrec$ for neutrino energy 1.2 GeV (top panel), 0.8 GeV (middle panel), and 0.4 GeV (bottom panel) calculated within the RFG model.}
\end{figure}
%%%%%%%%%%%%%%%%

To clarify this point, in Fig.~\ref{fig:givenE} we show a~comparison between the differential cross sections $d\sigma/dQ^2$ and $d\sigma/d\qrec$ obtained from the RFG model at
 {\em fixed} neutrino energies $E_\nu = $ 1.2 GeV, 0.8 GeV, and 0.4 GeV. With the exception of the lowest energy, the maxima are again in very good agreement. For this reason the cross sections averaged over a not-too-low-energy flux, like MiniBooNE's, are in good agreement.

 On the other hand, the
cross sections show quite a~rich structure, exhibiting bumps, dips, and knees, which make the $Q^2$ and
$\qrec$ distributions clearly different. While this structure is somewhat emphasized within
the RFG model, its origin is largely model independent, as it can be traced back to kinematics. In this context, the
most relevant feature is the location of the single particle strength, expressed by the average separation energy,
which turns out to be quite close in the RFG and SF models.

%{\bf This structure results from the simple momentum distribution of the RFG model and does not appear when the SF is used. %Thanks to its presence for the RFG model we are able to show in the following how the replacement of $Q^2$ with $\qrec$ r%edistributes the cross section. We want to emphasize that due to its kinematical mechanism the redistribution is model %independent; it is just easier to observe and to show in the case of the RFG model.}

In spite of the fact that the structure is not visible in Fig.~\ref{fig:Q2vsQ2rec}, as it is washed out by the flux average, the
results shown in Fig.~\ref{fig:givenE} imply that
neutrinos of fixed energy contribute in a~slightly different manner to $d\sigma/dQ^2$ and to $d\sigma/d\qrec$. At
neutrino energy 0.8 GeV, corresponding to the peak of the MiniBooNE flux, the difference between the two
cross sections is pronounced.

Before discussing why the replacement of $Q^2$ with $\qrec$ leads to a significant change of the cross section,
let us focus on the mechanism responsible for the observed structure in the case of $d\sigma/dQ^2$.
To find the value of $Q^2$ corresponding to the maximum we note that, due to Pauli blocking, the cross section may increase
with momentum transfer up to the value $\n{q}_*=2p_F$, needed to knock out nucleons sitting at the bottom of the Fermi sea.
The energy transfer can be written as
\be
\nonumber
\omega & = & E_\ve{p'} - E_\ve{p} + \EB  \\
& = &  \sqrt{M^2+(\ve p+\ve q)^2}-\sqrt{M^2+\ve p^2}+\EB,
\ee
where $M=(M_n + M_p)/2$, while $E_\ve{p}$ and $\ve p$ ($E_\ve{p'}$ and $\ve p'=\ve p+\ve q$) are the neutron (proton) energy and momentum. The above equation shows that, for any given
$\n p$, knocking out nucleons with momenta parallel to $\ve q$ requires the highest energy.
%It means that the minimal energy transfer being able to knock out any nucleon is equal to
As a consequence, the momentum and energy transfer $\n q_*$ and $\omega_*$, with
\beq
\omega_*=\sqrt{M^2+(3p_F)^2} - \sqrt{M^2+p_F^2} + \EB
\label{omaga_star}
\eeq
correspond to the maximum of the $Q^2$ distribution. The position of the maximum, $Q^2=0.146$ GeV$^2$, is
independent of neutrino energy, as long as $E_\nu$ is high enough for $\omega_*$ and $\n q_*$ to lie within the
kinematically allowed region, and not too close to its boundary. The bottom panel of Fig.~\ref{fig:givenE} shows
that for $E_\nu=0.4$~GeV this condition is not fulfilled.

%%%%%%%%%%%%%%%%
\begin{figure}
    \flushright
    \includegraphics[width=0.9\columnwidth]{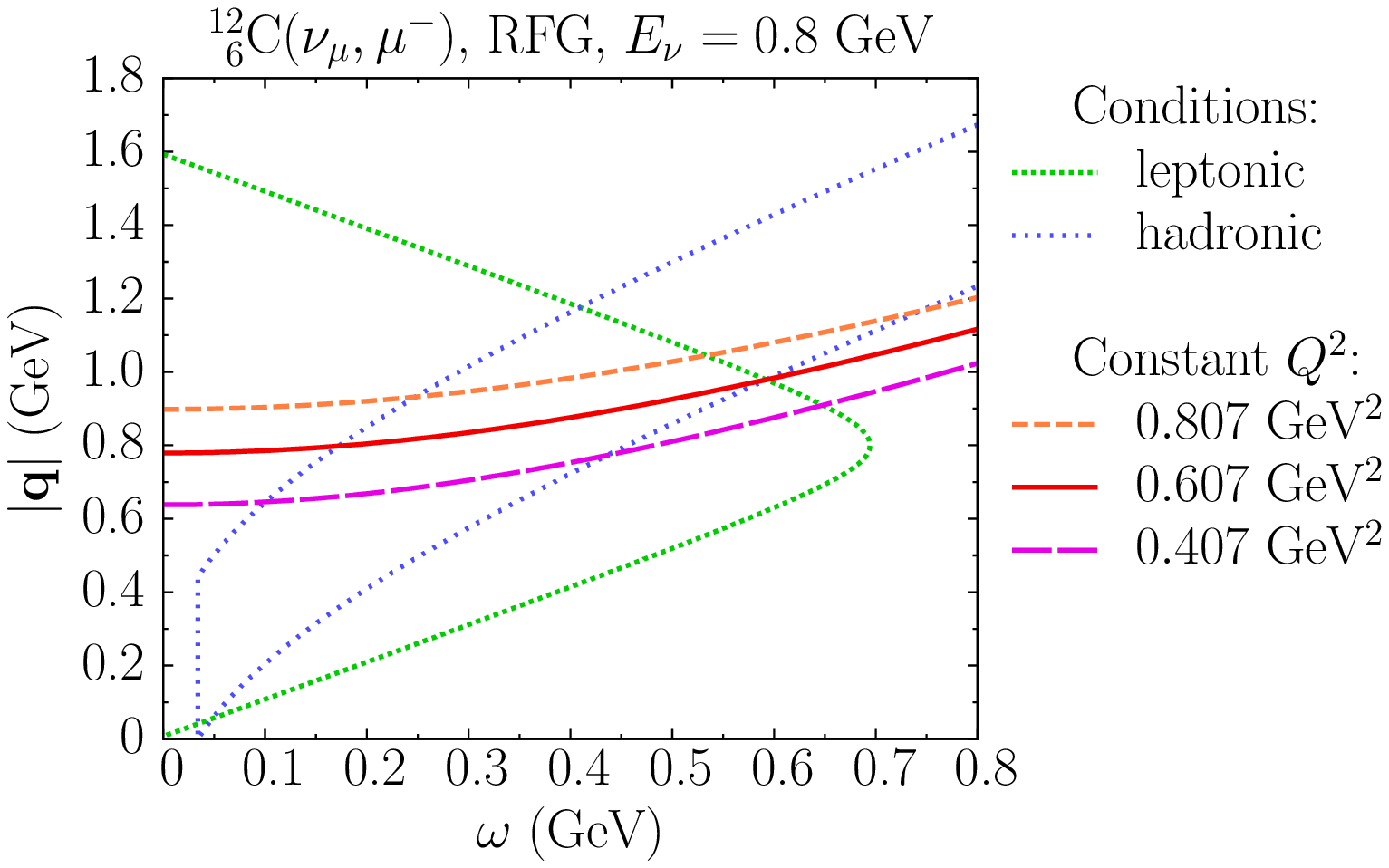}\\
    \vspace{1em}
    \includegraphics[width=0.9\columnwidth]{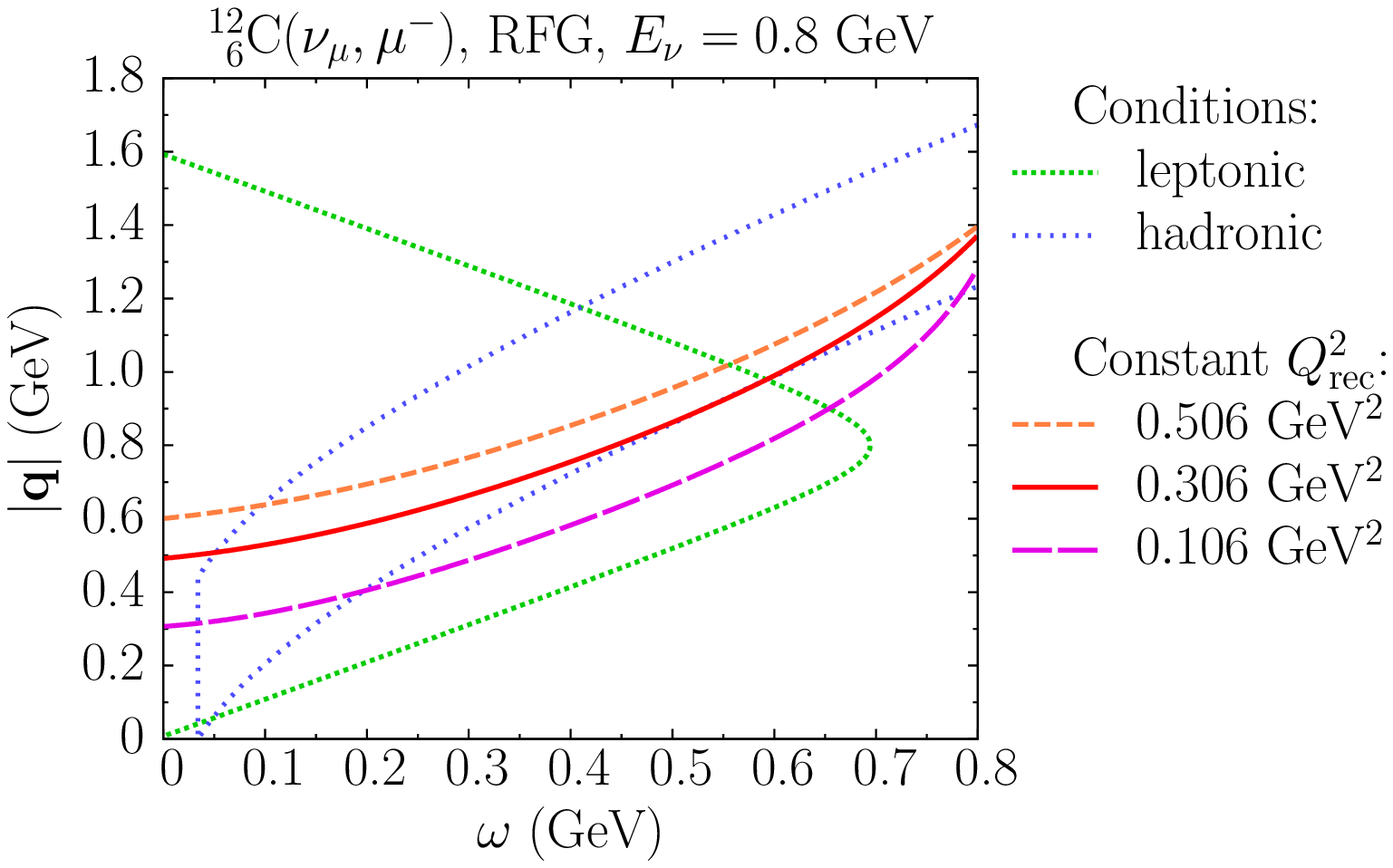}
\caption{\label{fig:phaseSpace} (color online). Upper panel: curves of fixed $Q^2$ in the $(\omega, \n q)$ plane, superimposed to the kinematically allowed region (within the dotted boundaries) for 0.8-GeV muon neutrino scattering. Lower panel: same as in upper panel but for fixed values of $\qrec$.}
\end{figure}
%%%%%%%%%%%%%%%%

The knee of the cross section, particularly visible for $E_\nu=0.8$ GeV, results from the phase space shrinkage
above a certain $Q^2$. The boundaries of the kinematically allowed region in the $(\omega,\n q)$ plane,
determined from the conditions
\beq\begin{split}
E_{\nu} -\n{k'}&\leq \n q\leq E_{\nu} +\n{k'},\\
\n{p'}-\n p&\leq \n q\leq \n{p'}+\n p,\\
0&\leq\n p\leq p_F,
\end{split}\eeq
where $\n{k'}=\sqrt{E_{\ve k'}^2-m_\mu^2}$ and $\n{p'}=\sqrt{E_{\ve p'}^2-M^2}$, are
\begin{equation}\begin{split}\label{eq:conditions}
h-p_F&\leq\n q\leq h+p_F,\\
E_\nu-l&\leq \n q\leq E_\nu+l,
\end{split}\end{equation}
with
\beq\begin{split}
h&=\sqrt{(E_F+\widetilde\omega)^2-M^2},\\
l&=\sqrt{(E_\nu-\omega)^2-m_\mu^2},\\
E_F&=\sqrt{M^2+p_F^2},\\
\widetilde\omega&=\omega-\EB.
\end{split}\eeq
As illustrated in the upper panel of Fig.~\ref{fig:phaseSpace}, where the kinematically allowed region for $E_\nu=0.8$~GeV lies within the dotted lines corresponding to the above conditions, the available phase space starts to decrease above $Q^2=0.607$ GeV$^2$. As a consequence, the differential cross section also decreases, exhibiting a~knee starting at this value of $Q^2$.
Comparison between the results corresponding to $E_\nu = 0.8$ GeV and $1.2$ GeV shows that as the neutrino energy
gets higher this effect becomes less significant. This pattern can be understood considering that for higher $E_\nu$ the reduction of the phase space starts at higher $\n q$ and $Q^2$, where the cross section is smaller. In addition, the shrinkage of the phase space is less pronounced, as the allowed $Q^2$ range is much broader.

The relation between $Q^2$ and $\qrec$ is by no means simple. For example, knowing only the energy loss~$\omega$ and the momentum transfer~$\n q$, in MC simulation we cannot determine $\qrec$; one additional independent quantity, e.g. the neutrino or muon energy, is necessary. This implies that using $\qrec$ we are not describing the intrinsic target response: some information on the interaction vertex is also involved. To calculate $\qrec$ we may use the equality
%%%%%%%%%%%%%%%
\begin{equation}\label{eq:E-kcosTheta}
E_\mu-\n{k'}\cos\theta=\frac{Q^2+m_\mu^2}{2E_\nu},
\end{equation}
%%%%%%%%%%%%%%%
following from the identity
\beq
k\cdot k'\equiv-\frac12\big[(k-k')^2-m_\mu^2\big],
\eeq
where $k=(E_\nu,\ve k)$ and $k'=(E_\mu,\ve k')$ are the neutrino and muon four-momenta, respectively. When we map $Q^2$ onto $\qrec$ using Eqs.~\eqref{eq:recQ2} and \eqref{eq:E-kcosTheta}, the resulting values are shifted as listed in Table~\ref{tab:knees}, which explains why the structures in $d\sigma/dQ^2$ and $d\sigma/\qrec$ appear at different positions. The lower panel of Fig.~\ref{fig:phaseSpace} illustrates the reason of the enhancement of the cross sections' knees, which turns them into bumps: where the phase space for $d\sigma/dQ^2$ shrinks, we observe an increase of the allowed range of $\qrec$;
the behavior in the two variables is thus completely different.

%%%%%%%%%%%%%%%%
\begin{table}
\caption{\label{tab:knees} Position of the knees or bumps of the distributions shown  in Fig. \ref{fig:givenE}.}
\begin{ruledtabular}
    \begin{tabular}{@{}ccc|cc@{}}
    $E_\nu$ (GeV)  &  $\omega$ (GeV)   & $\n q$ (GeV) &  $Q^2$ (GeV$^2$)  &  $\qrec$ (GeV$^2$)\\
    \hline
    0.4  & 0.2536  & 0.5013 & 0.187 & 0.107\\
    0.8  & 0.5923  & 0.9788 & 0.607 & 0.306\\
    1.2  & 0.9576  & 1.4181 & 1.094 & 0.494\\
    \end{tabular}
\end{ruledtabular}
\end{table}
%%%%%%%%%%%%%%%%

%%%%%%%%%%%%%%%%
\begin{figure}
    \flushright
    \includegraphics[width=1.0\columnwidth]{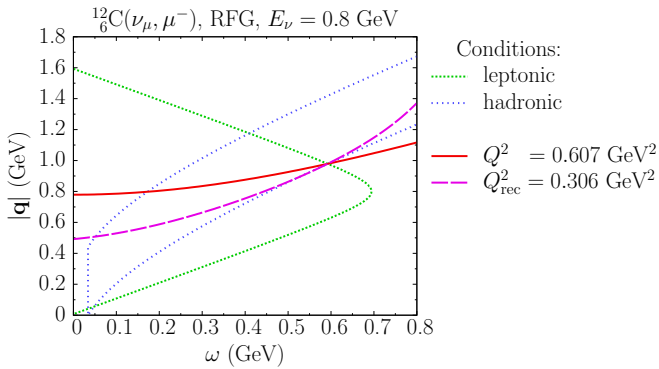}%
\caption{\label{fig:comp} (color online). Comparison of curves of fixed  $Q^2$ and $\qrec$ corresponding to the knee (bump) of the differential cross section $d\sigma/dQ^2$ ($d\sigma/\qrec$).}
\end{figure}
%%%%%%%%%%%%%%%%

%%%%%%%%%%%%%%%%%
%\begin{table}
%\caption{\label{tab:maxima} Position of maxima of the cross section for selected values of energy.}
%    \begin{tabular}{@{}c|cc@{}}
%    $E_\nu$ (GeV)  &  $Q^2$ (GeV$^2$)  &  $\qrec$ (GeV$^2$)\\
%    \hline
%    0.4  & 0.146 & 0.094\\
%    0.8  & 0.146 & 0.124\\
%    1.2  & 0.146 & 0.132\\
%    \end{tabular}
%\end{table}
%%%%%%%%%%%%%%%%%

In Fig.~\ref{fig:comp} we compare the lines of constant $Q^2=0.607$ GeV$^2$ and $\qrec=0.306$ GeV$^2$ corresponding
to $E_\nu = $ 0.8 GeV. The results clearly show that the two variables, in spite of leading to very similar distributions of events,
are quite different. It clearly appears that the $Q^2$ and $\qrec$ distributions are determined by
the nuclear response in different regions of the $( \omega,\n q)$ plane, the latter being sensitive to significantly
lower values of the momentum transfer. This feature implies that the $\qrec$ distribution receives contributions
from the kinematical region in which the validity of the IA may become questionable.

\section{Breakdown of the impulse approximation}
\label{lowq}

In order to pin down the boundary of the kinematical region in which the IA is applicable, we have
studied the response of nuclear matter (a translationally invariant system consisting
of equal number of protons and neutrons subject to strong interactions only)
to a scalar probe delivering momentum ${\bf q}$ and energy $\omega$:
\beq
S({\bf q},\omega) = \sum_n  \langle 0 | \rho^\dagger_{{\bf q}} | n \rangle
\langle n | \rho_{{\bf q}} | 0 \rangle\: \delta(\omega + E_0 - E_n ).
\label{def:resp}
\eeq
In the above equation, the operator $\rho_{{\bf q}}= \sum_{{\bf p}} a^\dagger_{{\bf p}+{\bf q}} a_{{\bf p}}$,  $a^\dagger_{{\bf p}}$ and $a_{{\bf p}}$ being nucleon creation and annihilation operators,
describes the fluctuations of the target density induced by the
interaction with the probe. The target ground and final states, $\vert 0 \rangle$ and $\vert n \rangle$,
are eigenstates of the nuclear Hamiltonian $H$ belonging to the eigenvalues $E_0$ and $E_n$,
respectively. Note that scattering cross section of a scalar probe (e.g. thermal neutrons
scattering off liquid helium) can be written in the simple form
\beq
\frac{d \sigma }{d \Omega} = \left( \frac{d \sigma }{d \Omega} \right)_0 \ S({\bf q},\omega) ,
\eeq
where $(d \sigma / d \Omega)_0$ is the elementary cross section describing scattering off individual target
constituents, while the response defined in Eq.~\eqref{def:resp} is an {\em intrinsic} property, fully determined
by internal target dynamics.

In the case of electron- or neutrino-nucleus scattering, Eq.~\eqref{def:resp} can be readily generalized,
replacing the density fluctuation operator with the operators describing the electromagnetic and weak
nuclear currents. The resulting response tensor reads
\beq
W_{\nu\mu} = \sum_n  \langle 0 | J_\nu^\dagger | n \rangle
\langle n | J_\mu | 0 \rangle\: \delta^4(q+ p_0 - p_n ).
\eeq

In the IA regime, Eq.~\eqref{def:resp} can be rewritten in the form
\beq
\label{L0}
S({\bf q},\omega) = \int d^3p\, dE\: P_h({\bf p},E)\: P_p({\bf p}+{\bf q},\omega-E),
\eeq
where the hole and particle spectral functions, $P_h$ and $P_p$, describe the energy and momentum
distributions of the struck particle in the initial and final states, respectively. The simplest implementation
of the IA (usually referred to as plane wave impulse approximation, or PWIA) is based on the further
assumption that final state interactions (FSI) between the knocked out nucleon
and the spectator particles can be neglected. As a consequence, the nuclear matter particle spectral
function reduces to
\be
P_p(\ve p +\ve q,\omega-E) & = & [ 1 - \frac{4}{3} \pi p_F^3 \ n(\ve p +\ve q)]\nonumber \\
& \times &  \delta( \omega-E+M - E_{\ve{p}+\ve{ q}}),
%\beq
%P_p(\ve p +\ve q,\omega-E)  = n(\ve p +\ve q) \  \delta( \omega-E+M - E_{\ve{p}+\ve{ q}}) \ ,
\label{part:sf}
%\eeq
\ee
where $n(\ve p +\ve q)$ is the occupation probability of the single particle state of momentum
$\ve p +\ve q$ and $E_{\ve{p}}$ denotes the energy of a free nucleon carrying momentum $\ve{p}$.
Note that, in the above definition, Pauli blocking of the phase space available to the struck nucleon is
treated in a consistent fashion, using the momentum distribution obtained from the hole spectral
function through
\beq
n({\bf p}) = \int dE\: P_h({\bf p},E).
\eeq
In most calculations of electron- and neutrino-nucleus scattering cross sections, $ 4 \pi p_F^3 n({\bf p})/3$ is replaced with
the RFG result $\theta(p_F - |{\bf p}|)$, yielding
\beq\begin{split}
P_p(\ve{p'},{\cal E}) &= \big[1 - \theta(p_F-\n{ p'})\big] \, \delta( {\cal E}+M - E_{\ve{p'}})\\
&=\theta(\n{ p'}-p_F)\, \delta( {\cal E}+M - E_{\ve{p'}}) \ .
\end{split}\eeq
It has to be pointed out that the above prescriptions, while being justified in the case of uniform nuclear matter,
are questionable when applied to nuclei. In nuclear matter, due to translation invariance, the linear momentum
is a good quantum number, that can be used to label single particle states. As a consequence, the momentum
distribution  also provides the occupation probability of the states. In finite nuclei, on the other hand, single particle
states must be labeled according to the total angular momentum ${\bf J}$. In this case, for any given ${\bf p}$, $ n({\bf p})$
receives contributions from states of different ${\bf J}$, and may even exceed unity.

The available results of accurate nuclear matter calculations can be used to model the particle spectral
function of finite nuclei within the framework of the local density approximation \cite{SFCarbon}, i.e.
using the definition of Eq.~\eqref{part:sf} with
%\beq
%n({\bf p}) = \int d^3r\  n_{NM}[\rho(r),{\bf p}]\ \rho_A(r) \ ,
%\label{np:LDA}
%\eeq
\beq
\frac{4}{3}\pi p_F^3 n({\bf p}) \rightarrow \int d^3r\: \frac{4}{3} \pi p_F^3\:
n_{NM}[\rho(r),{\bf p}]\:\rho_A(r) \ ,
 \label{np:LDA}
\eeq
where $\rho_A(r)$ is the nuclear density distribution, normalized to unity, and $n_{NM}[\rho(r),{\bf p}]$ is the momentum distribution of
nuclear matter at uniform density $\rho(r)$. This procedure has been used in all calculations of nuclear cross
sections discussed in this paper.

Within the nonrelativistic approximation, in which both the response and the hole spectral function can
be evaluated using realistic nuclear Hamiltonians, the validity of the IA can be tested comparing
$S({\bf q},\omega)$ of Eq.~\eqref{def:resp} to
\be
\nonumber
\label{PWIA}
S_\textrm{PWIA}({\bf q},\omega) & = & \int d^3p \: dE\: P_h({\bf p},E)  \left[1 - \frac{4 \pi}{3} p_F^3\: n({\bf p}+{\bf q})\right] \\
 & &\quad\times\:\delta \left( \omega-E-\frac{|{\bf p}+{\bf q}|^2}{2M} \right),
\ee
for different values of the momentum transfer ${\bf q}$.

The nuclear matter $S({\bf q},\omega)$ at equilibrium density, $\rho_0 = 0.16$ fm$^{-3}$
(corresponding to $p_F = 262.4$ MeV), has been recently
computed using the correlated basis function formalism and an effective interaction derived
from a state-of-the-art para{\-}metrization of the nucleon-nucleon potential~\cite{BF09}. To analyze the interplay
between short- and long-range correlations,  the response defined as in Eq.~\eqref{def:resp} has been evaluated in
both the Hartree-Fock and Tamm-Dancoff approximations.

\begin{figure}
\begin{center}
\includegraphics[scale=0.78]{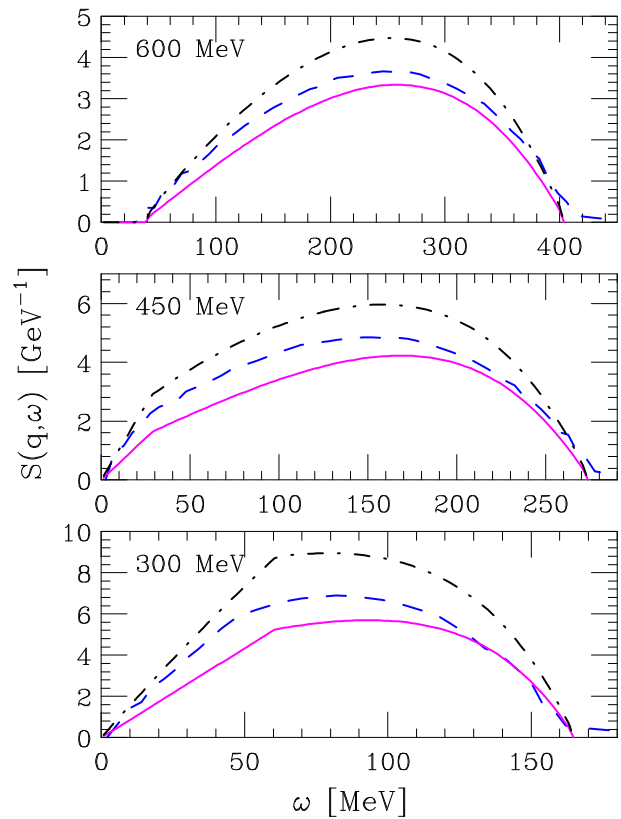}
\end{center}
\caption{\label{comp:sqw}Energy dependence of the nuclear matter response $S({\bf q},\omega$). Solid lines:
correlated Hartree-Fock approximation~\cite{BF09}. Dashed lines: PWIA results, obtained from
  Eq.~\eqref{PWIA} using the SF of Ref.~\cite{BFF89}. Dot-dashed lines: results of the
  Fermi gas model (with nonrelativistic energies) at $k_F = 262.4$ MeV, corresponding to the
  equilibrium density of nuclear matter. The panels are labeled
according to the value of $\n{q}$.}
\end{figure}

Figure~\ref{comp:sqw} shows a comparison between the responses of Ref.~\cite{BF09}, obtained using
Eq.~\eqref{def:resp} and the correlated Hartree-Fock approximation, and those obtained from Eq.~\eqref{PWIA}
using the nuclear matter hole spectral function of Ref.~\cite{BFF89}. The main difference between the
two calculations lies in the treatment of the target final state. In the IA scheme the state describing the struck
particle is factored out, while in the approach of Ref.~\cite{BF09} the final $A$-nucleon state includes
both statistical and dynamical correlations between the struck particle and the spectators.To make the comparison fully consistent, the PWIA response has been computed including only
the contributions of one-hole final states to $P_h({\bf p},E)$. The results of the Fermi gas model
with nonrelativistic kinetic energy spectrum
%at the equilibrium density of nuclear matter,
are also displayed.

%%%%%%%%%%%%%%%%%%%%%%%%%%%%%%%%%%%%%%%%%%%%%%%%%%%%%%%%%%%%%%%%%%
%%%%%%%%%%%%%%%%%%%%%%%%%%%%%%%%%%%%%%%%%%%%%%%%%%%%%%%%%%%%%%%%%%

The results of Fig.~\ref{comp:sqw} clearly show that at $\n q < 2p_F$ the response obtained
from Eq.~\eqref{PWIA} does not exhibit the linear behavior at low $\omega$ resulting from the
antisymmetrization of the final state. On the other hand, as the momentum transfer increases, the PWIA response
draws closer to the one obtained in Ref.~\cite{BF09}. At $\n q \sim 600$ MeV the results of the two approaches are
within 10\% of one another in the region of the maximum. Note that inclusion of dynamical FSI, e.g. according to the
approach of Ref.~\cite{gofsix}, would produce a quenching of the PWIA response in the top panel of Fig. \ref{comp:sqw},
thus bringing the solid and dashed lines in even better agreement. Theoretical studies of electron-nucleus scattering~\cite{ref:BFNSS}
suggest that at $\n q$ $\sim$ 600 MeV, the FSI effect at the quasifree peak is $\sim$10~\%.

The emerging pattern suggests that the assumptions
underlying the IA are likely to be valid at momenta larger than $\sim$$2p_F$, while at lower $\n{q}$ factorization
does not appear to provide an adequate description of the final state. In addition, it has to be kept in mind that
at $\n q \sim p_F$ long-range correlations,
involving more than one nucleon, also play a significant role~\cite{BF09}.

Despite the fact that, being based on the nonrelativistic approximation, the approach of Ref.~\cite{BF09} should not be used in the calculation of the MiniBooNE event distribution, it helps to clarify why the $\qrec$ distribution of events measured by this experiment can be described by the RFG model~\cite{ref:MiniB_eff} and the SF approach only for $\qrec\geq0.25$~GeV$^2$.

To improve the description at lower $Q^2$, the MiniBooNE Collaboration introduced the {\em ad hoc} additional parameter
$\kappa$ in the RFG model. While the authors of Ref. \cite{ref:MiniB_eff} argue that this procedure allows for
a better treatment of Pauli blocking, we think that the inclusion of $\kappa$ cannot be justified on physics ground. Therefore,
we refer to the modified RFG model of Ref. \cite{ref:MiniB_eff} as the $\kappa$-fit.

\begin{figure}
    \centering
    \includegraphics[width=0.80\columnwidth]{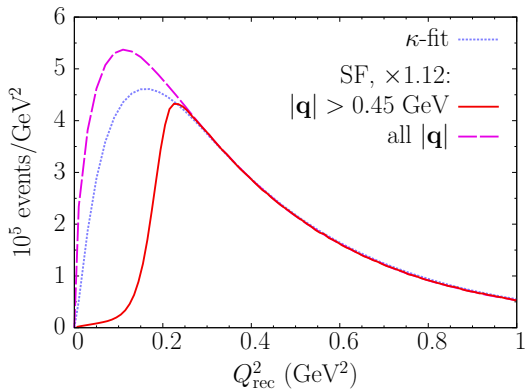}%
\caption{\label{fig:low_q} (color online). Comparison of the MiniBooNE parametrization of the data (dotted line), labeled as the $\kappa$-fit, to the spectral function calculation (dashed line). The solid line depicts the contribution to the latter from the region where the IA is expected to be valid. The SF results are multiplied by a factor 1.12 to make them match the $\kappa$-fit.}
\end{figure}

Figure~\ref{fig:low_q} shows that in the region of lower values of $\qrec$ the contribution of $\n q$'s below $2p_F$ is dominant. Hence, to explain the neutrino-nucleus cross section $d\sigma/d\qrec$ at quantitative level in the whole range of $\qrec$ one needs a consistent description of both short- and long-range correlations.
%in other words scattering off both single and multiple nucleons bound in the nucleus.
Unfortunately, due to the difficulties involved in the treatment of relativistic particles in the final state,
%To the best of our knowledge
such a description has not been fully developed yet.

%%%%%%%%%%%%%%%%
%%%%%%%%%%%%%%%%

\section{Experimental issues}
\label{expt:issues}
The average axial mass extracted by past experiments turns out to be significantly lower than the values recently reported
 by K2K and MiniBooNE~\cite{ref:Kuzmin}. Although the early experiments were carried out using a variety of targets,
about 60\% of the total number of recorded events were collected on deuteron. The results of deuteron measurement
with highest statistics \cite{ref:BNL90,ref:ANL82,ref:BNL81} are very consistent with one another and have small error bars.
Compared to deuteron experiments,
those carried out with heavier targets have much poorer statistics, the typical number of events being lower by one
order of magnitude. Moreover, using a deuteron target allows one to minimize
the systematic error associated with the treatment of nuclear effects.

A lower value of the axial mass is also supported by the result recently obtained by the NOMAD Collaboration~\cite{ref:NOMAD}
using a carbon target.

On the other hand, MiniBooNE collected more events than all other experiments combined, and carried out an
analysis based on the shape of the reconstructed $Q^2$ distribution.
MiniBooNE reported 193709 events surviving the cuts, of which about 180000 correspond
to  $0 \leq \qrec < 1$ GeV$^2$. In the analysis not involving the additional parameter $\kappa$, two cuts were applied:
one at low $\qrec$, to exclude the region
where the RFG model is expected to break down,  and one at high $\qrec$, to exclude the region of low statistics.
After rejection of the events excluded by the cuts, the data sample
employed to extract the axial mass reduced to $\sim$112000 events, i.e. $\sim$62\% of the total. Note, however,
that this figure is still $\sim$60 times larger than the number of events typically collected in deuteron
experiments.

\subsection{Identification of CCQE events}
\label{identification}

The first problem to be addressed in the comparison between theoretical calculations and experimental
results is background simulation, which is, to a significant extent, detector dependent.

The extraction of the axial mass requires an accurate selection of the CCQE events and a quantitative description of the
 irreducible backgrounds which may change the topology of the observed event.
 For example, a $\Delta$-production process followed by pion absorption may be undistinguishable from a
 CCQE interaction, whereas a primary CCQE process may yield pions due to final state interactions.
Therefore, a~more complete theoretical analysis should account for intranuclear cascade in a consistent way.

Although
cascade calculations including nucleon-nucleon correlations have not been developed yet, we can gauge the relevance
of these effects using the results of a comparison of the MC generators employed in the analysis of neutrino experiments,
reported in Ref.~\cite{ref:LadekMC}. The predicted ratio between {\em true} and {\em observed} CCQE events
turns out to be in the range 82\%--89\% for a 1-GeV $\nu_\mu$ beam and oxygen target. On the other hand, the fraction
of events misidentified as inelastic due to pion produced in final state interactions is typically 0.5\%--3\%. The uncertainty in the modeling
of pion production and absorption may be partly responsible for the disagreement between the values of the axial mass reported by K2K and MiniBooNE and those resulting from different experiments.

Scattering off a~correlated pair of nucleons may also be misinterpreted as pion production, as it produces two hadron tracks
in the final state. However, even neglecting nucleon absorption this background does not exceed 3\% of the CCQE events in
the MiniBooNE kinematical range, and is therefore not likely to be significant.
Because of the weaker $Q^2$ dependence, resulting from the higher nucleon removal energy, a proper interpretation of
these events may only marginally increase the value of axial mass extracted from the analysis.

The effect of FSI, negligible in light nuclei and not taken into account in this paper, is expected to make the $\qrec$ distribution flatter, quenching its maximum and redistributing strength towards higher values of the four-momentum transfer~\cite{ref:BFNSS}.
This behavior is due to the fact that FSI couple one-hole states to one-particle--two-hole states, thus leading to an increase of the average removal energy. As a result, the neutrino-nucleon interaction takes place at higher $Q^2$. Hence,
part of the discrepancy between the results of K2K and MiniBooNE, on the one hand, and deuteron-based experiments, on the other hand, may be ascribable to FSI effects.

It is also very important to realize that the influence of reaction mechanisms beyond the IA cannot be minimized increasing the beam energy~\cite{ref:IA}. It turns out that low momentum transfers ($\n q\lsim p_F$) provide almost the same contributions to
the CCQE cross section for neutrino energy 0.4 GeV and 100 GeV. In the absence of a fully consistent theoretical description
of all the relevant mechanisms, the most reasonable option for the experimental analysis appears to be
a~cutoff, to reject events with $\n q\lsim2p_F$.

It should be kept in mind that in the region where the IA is not valid the cross section may not scale with the number of
nucleons. As a consequence,  the axial mass (or any other parametrization of the axial form factor) extracted from
neutrino-nucleus data at low $\qrec$ may turn out to be target dependent.

\subsection{Proposal of new kinematical variables}
\label{newvariable}

The axial mass is currently extracted from experimental data using the shape of the $\qrec$ distribution of CCQE events.
However, as pointed out in Section \ref{sec:Q2}, $\qrec$ cannot be directly measured; its definition [see Eq.~\eqref{eq:recE}] involves the nucleon
separation energy and depends on the applied approximations, e.g. the assumption that the struck nucleon be at rest.
%like neglecting the difference between nucleon masses,
%Besides, the reconstruction works a~bit worse for the spectral function.

%%%%%%%%%%%%%%%%
\begin{figure}
    \centering
\includegraphics[width=0.80\columnwidth]{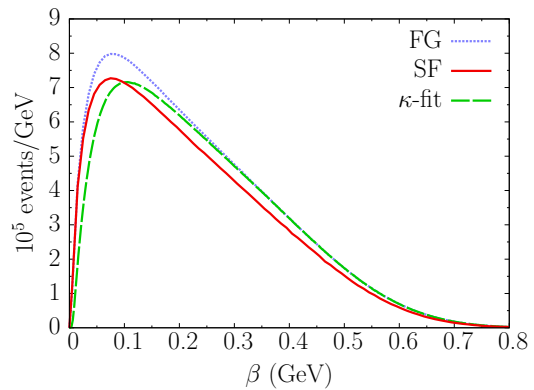}%
%    \vspace{1em}
%\includegraphics[width=0.80\columnwidth]{beta_shape}%
\caption{\label{fig:beta_abs&shape} (color online). Distribution of the MiniBooNE data calculated using the ``$\kappa$ fit'' (dashed line), the RFG model (dotted line), and the SF approach (solid line) as a function of  the variable $\beta$, defined in
Eq.~\eqref{def:beta}.
%The upper panel shows the predicted dependence of the absolute number of events on $\beta$, the lower panel
%compares the shapes of the distributions (the SF results are multiplied by 1.12).}
}
\end{figure}
%%%%%%%%%%%%%%%%

It may be convenient replacing $\qrec$ with the new, model independent, variable
%%%%%%%%%%%%%%%
\begin{equation}
\label{def:beta}
\beta=E_\mu-\n{k'}\cos\theta,
\end{equation}
%%%%%%%%%%%%%%%
whose definition only involves measured quantities. From
\beq
\beta=\frac{k\cdot k'}{E_\nu}=\frac{Q^2+m_\mu^2}{2E_\nu},
\eeq
it follows that the $\beta$ distribution exhibits the same behavior as the $\qrec$ distribution, and can be comparably
useful in the extraction of the axial mass or in the analysis of nuclear effects. Figure~\ref{fig:beta_abs&shape} shows
the distributions obtained for the MiniBooNE flux as a function of $\beta$.

An even better choice appears to be provided by the variable
\beq
\phi = \frac{1}{m_\mu + \beta}\:.
\label{def:phi}
\eeq

\begin{figure}
    \centering
% \includegraphics[width=0.80\columnwidth]{beta_abs}\\%
%    \vspace{1em}
   \includegraphics[width=0.80\columnwidth]{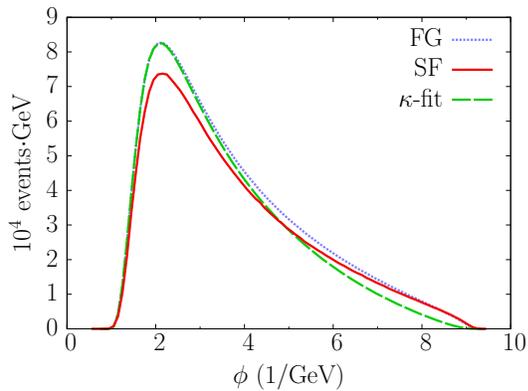}%
\caption{ \label{phi:dist} (color online). Same as in Fig.~\ref{fig:beta_abs&shape}, but plotted as a
function of the variable $\phi$, defined in Eq.~\eqref{def:phi}.}
\end{figure}

Figure~\ref{phi:dist}, showing the $\phi$ distributions of the MiniBooNE data obtained using different approaches,
clearly illustrates that the main advantage of using $\phi$ lies in the fact that
the deviations of the $\kappa$ fit from the RFG model, reflecting the breakdown of the IA, show up in the
high $\phi$ tail of the distribution. As a consequence, a single cut
at $\phi = 3$ GeV$^{-1}$ allows one to reject both the region of low statistics and the region where
effects beyond the IA are expected to be important.
Hence, the extraction of the axial mass from the $\phi$ distribution would be based on a larger data sample.
Moreover, the dependence of the $\phi$ distribution on the axial mass turns out to be only visible
at $\phi >$ 1.6 GeV$^{-1}$, corresponding to the region of highest statistics. Note that in the case of the
$\qrec$ distribution the situation is reversed.

%%%%%%%%%%%%%%%%
%%%%%%%%%%%%%%%%
\section{conclusions}
\label{concl}

In this paper, we have analyzed possible sources of the observed disagreement between the
$\qrec$ distributions of CCQE events reported by several recent experiments and the prediction
of Monte Carlo simulations.

As far as the treatment of nuclear effects is concerned, our work suggests that, in addition to the known
limitations of the RFG model, discussed in, e.g., Refs. \cite{ref:BFNSS,BM},  the most critical
feature of present analyses is the assumption that the IA scheme be applicable over the whole range of $Q^2$.
Electron scattering studies have provided ample evidence that the IA breaks down in the region of
low momentum transfer, which turns out to provide a significant fraction of the observed CCQE events,
independent of neutrino energy.

The failure of the IA is clearly exposed by the comparison between the full nuclear matter response and
the IA result, discussed in Section \ref{lowq}, showing that at momentum transfers $|{\bf q}| <  2 p_F$ the contributions
of more complex reaction mechanisms become important, or even dominant.

While our calculations focus on the effects of factorization of the final state, it must be pointed out
that different mechanisms should also be considered. For example, meson exchange currents, which are long
known to provide appreciable contributions to the electron-nucleus cross section at the
quasi elastic peak and beyond, are also expected to contribute to the
background in the case of neutrino-nucleus scattering.

The development of a consistent treatment of scattering processes at low and high momentum
transfer within a formalism easily implementable in MC simulation, while being feasible, involves severe difficulties,
and will require a significant effort in the years to come. On the other hand, we believe that introducing {\em ad hoc}
modifications of the available models, lacking a sound physical interpretation, will not help to clarify the
origin of the disagreement between theoretical predictions and observations.

Among the issues related to data analysis, identification of CCQE processes appears to be prominent.
The discussion of Section~\ref{identification}, based on the results of MC simulations, indicates that
it may be at least partly responsible for the disagreement between the values of the axial mass recently reported by
K2K and MiniBooNE and those obtained from different measurements.

Improving event identification will require a more realistic description of FSI, combining the intranuclear cascade approach
with a fully realistic description of the target nucleus, and including the relevant inelastic channels leading to pion
production. The key elements needed to pursue this project, i.e. in-medium nucleon and pion cross sections and nuclear
wave functions including correlation effects, can be extracted from the available data and from theoretical results of accurate many-body calculations.

Finally, we suggest that data analysis might be improved using a new kinematical variable which, unlike $\qrec$, can
be defined in terms of measured quantities only. In addition, using the new variable would allow one to reject both
the region of low statistics and the region where effects beyond the IA are expected to be important with a single cut.
As a result, the extraction of the axial mass would be based on higher event statistics.

\begin{acknowledgments}
A.M.A. was supported by the Polish Ministry of Science and Higher Education under Grant No. 550/MOB/2009/0. O.B. and N.F. were supported by INFN under Grant PI31.
\end{acknowledgments}

\end{document}